\documentstyle[12pt]{article}
\evensidemargin\oddsidemargin
\begin{document}
\count0 = 1
\begin{titlepage}
\vspace {20mm}
\begin{center}
\ QUANTUM SPACE-TIME TRANSFORMATIONS\
\smallskip
\ AND REFERENCE FRAMES STATES\\
\vspace{6mm}

\bf{S.N. MAYBUROV}
\vspace{6mm}

\small{Lebedev Institute of Physics}\\
\small{Leninsky pr. 53, Moscow Russia, 117924}\\

\vspace{15mm}
\small{\bf Abstract}
\end {center}
\vspace{3mm}

\small{
 We argue that correct account of the quantum properties of macroscopic
 objects which form reference frames (RF) demand
 the change of the standard space-time picture accepted in
 Quantum Mechanics. Galilean or Lorentz space-time transformations
 are shown to become incorrect in this case and for the
 description of transformations between different RF
 the special quantum space-time transformations are introduced.
 Consequently it results in the generalised Schrodinger equation
 which depends on the observer mass.
The experiments with macroscopic coherent states are proposed
in which this effects can be tested.}

\vspace{14mm}
\small {Proceedings of 6th Quantum Gravity Seminar, Moscow ,
 (World Scientific,Singapore, 1996) }\\
\vspace {18mm}
\small {---------}\\
\small {  * E-mail  Mayburov@sgi.lpi.msk.su}
\end{titlepage}
\begin{sloppypar}
\section{\large{Introduction}}

In a modern Quantum Mechanics (QM) the particles states and other objects
 evolve in Minkowski space-time regarded as independently existing
 entity. Alternatively it's
 the main method of the description of the surrounding  world from the point
 of  view of particular observer. In Classical Physics it corresponds to
 introduction of space coordinate axes associated with particular
 reference frame(RF) which is supposed to be some solid macroscopic object
 of nonzero mass
 or the system of them. Despite that in QM
 the behavior of physical objects can be strikingly nonclassical it's
 tacitly assumed describing RF properties in QM
 that any RF evolution is always exactly classical.
 Consequently all RF coordinate transformations in QM are supposed to be
 identical to classical - Galilean or Lorentzian ones.
 In our paper we argue that this assumption is in general incorrect and
 quantum features of RF should result in a special quantum space-time
 transformations. We must stress that this results will be obtained without
introducing new axioms or hypothesys ,but staying in the framework of
 standard QM. In our opinion current discussions of Quantum space-time
should include the correct definition of quantum RF $\cite{Dop}$
First the importance of RF quantum properties was noticed in
Quantum Gravity study of distributed - dust or fluid RF
$\cite{Rov},\cite{Kuc}$. Yet their detailed analysis even for flat
 space-time concerned with Quantum Measurements aspects wasn't
performed up to now, only some phenomenological discussions were
published $\cite{Bacry}$.

In this paper only one quantum  RF effect
 will be analysed. Namely it's the predicted by QM
 existence of the wave packet of the macroscopic object defined as RF,
 which gradually enlarge with time. Despite that its
 scale in the standard laboratory conditions is quite small ,we obtain
that it can have important meaning both for Cosmology and for small
(up to Plank scale) distance Physics.  It'll be shown that in nonrelativistic
 QM  quantum RF transformations corresponds to the additional quantum
symmetry.
 We descibe also the special experiments which can be proposed to test this
 conclusions  using modern experimental technics.

 The formulated problem is closely related
 to the  macroscopic quantum
 coherence topics  which embrace different observations of the superpositions
 of the macroscopic objects states. The recent studies
 have shown that for low dissipation superconductor systems
the superpositions of the macroscopic states can be observed
 $\cite {Leg}$. The experimental tests of this effects
with SQUID rings are now prepared $\cite{Prance}$.

In distinction from Classical Physics
 in QM framework the system defined as RF presumely should be able to measure
the obsevables of quantum states i.e. to be quantum observer.
 At first sight it seems this problem can be
solved only when the detailed microscopic model of state vector collapse
 will be developed. Despite multiple proposals up to now well established
theory of collapse which answer all difficult questions is absent $\cite{May}$.
  Alternatively we'll show that our problem premises doesn't connected
 directly with the state vector collapse description
$\cite{Wigner}$. In place of it we'll make two simple assumptions
on the observer system properties , which are in the same time rather weak.
  The first one is that the observer system (OS) or RF consists of the
 finite number of atoms and have the finite mass.
It was argued recently that even interaction with the single
 atom can result in the collapse of particle wave function , which seems
to us quite sensible $\cite{Van}$.

  So in this paper we assume that
observer system (OS) is the ensemble of microparicle detectors , meters and
 recording
devices which perform the coordinate or other measurements on quantum
 objects. As the realistic example we can regard photoemulsion plate or
diamond crystall which can measure microparticle position relative to its
c.m. and simultaneously record it.
 We'll regard measuring device to be in a
 pure state as usually made in a Measurement Theory.

 The experiments with the atomic
 and molecular beams confirm
 that complex quantum system can be obtained in delocalised state
without change of its internal properties.
  We'll consider the hypothetical
situation when the free observer $O_1$ is described by some other
macroscopic observer $O_M$ as a pure quantum state with large
uncertainty of centre of
mass coordinate $R_c=\sum m_i*r_i/M$.
The question which we had in mind preparing this work was : if the observer
 can be in such
delocalised quantum state what will he see looking at the objects of
 our macroscopic world ?

 It's well known that solution of Shrodinger equation for
  any free quantum system in a pure state consisting of $N$
constituents can be presented as the
\begin {equation} 
  \Psi(r_1,...,r_n,t)=\sum c_l\Phi^c_l(R_c,t)*\phi_l(r_{i,j},t)
\end {equation}
where   $r_{i,j}=r_i-r_j$  are  internal coordinates
 of constituents
 $\cite{Schiff}$. Here $\Phi^c_l$ describes the c.m. motion of the system.
 It demonstrates that in QM framework the functioning  and evolution of the
 system in the absence of the external fields is separated into the external
evolution as whole of the pointlike particle M and internal evolution
 completely defined by $\phi_i(r_{i,j},t)$ if constituents
 interaction
depends only on $r_{ij}$ as usually have place. So the internal evolution is
 independent of whether the  system is localised
 in macroscopic reference frame (MRF)
 or not. Relativistic QM and Field Theory studies show that the
 factorisation of c.m.
 and relative motion holds true even for nonpotential forces and
 variable $N$ in secondary quantised systems $\cite{Shw}$.
 Moreover this factorisation expected to be correct for nonrelativistic
 systems
 where binding energy is much less then its mass $m_1$, which is
 characteristic for most of real detectors. For our problem
 it's enough to assume
   the factorisation of c.m. motion holds for the observer system
  only in the time interval $T$
 from the system preparation procedure , until the act of measurement starts
,i.e. when measured particle collides with it. More exactly our second and
 last assumption about observer properties is
that the during period $T$ its state is decribed by the wave function
generalising  (1) of the form
$$
 \Psi(R_c,q,t)=\sum c_l\Phi^c_l(R_c,t)*\phi_l(q,t)
$$
 where $q$ denote all internal detector degrees of
freedom which evolve during $T$ according to Schrodinger equation. To
 simplify our calculations
we'll take below all $c_j=0$ except $c_1$ which wouldn't influence our final
 results.

   Any free object
  which was initially localised in a wave packet $\Phi(R,t_0)$
 after it will gradually smear in space at a low speed
,which for gaussian packet with initial dispersion $a_0$ described
as $\cite{Schiff}$ :
\begin {equation}    
a(t)=(\int(R-\bar{R})^2|\Phi(R,t)|^2d^3R)^{1/2}
 =a_0(1+\frac{t^2}{m^2a_0^4})^{1/2}
\end {equation}
(Plank constant $\hbar$ and $c$ in our calculations is equal $1$ ).
The standard conclusion is that to observe experimentally
measurable smearing of macroscopic object
demands too large time , but we'll show that for some mesascopic experiments
it can be reasonably small to be tested in the laboratory conditions.
In our work it's  permitted  ad hoc preparation of any initial
 state vector described by the smooth
function $\Psi(r_i,t_0)$ in agreement with QM postulates.
 During this study we
assume that RF and OS are always identical entities, but we'll discuss their
possible distinctions in conclusion which in fact can weaken demands to RF
formulated in this chapter.

\section{\large{Q-transformations Formalism}}

To explain our approach to RF transformations in QM in a simpliest terms
we consider gedankenexperiment (GEO) in which wave packet of observer system
 $O$ can be studied.
 Its layout where gravitation force is absent includes
 collimated along the Z axe neutron source $S_n$
 installed inside vacuum chamber and $O$ suspended in its volume
  without any contact with the walls . $S_n$ can
 emit one by one
 well-timed neutrons with mass $m_2$ in a very narrow beam with coordinates
 $x_s$ ,$y_s$, so that their wave function $\psi_n(x_n)$ can be approximated
 by the delta-function $\delta(x_n-x_s)$ . All states are considered at
fixed time $t_0$ and due to $t$ dependence
in $\psi$ arguments omitted , until evolution isn't become essential.
  For the simplicity the wave packet
 of the free observer $O$ is supposed to
 smear  significantly only along X axe and described by the
 wave function $ \psi_1(x_1)$ (internal wave functions can be neglected
 as we argued).
All this wave functions are defined in MRF connected with some
macroscopic object $M_R$ which mass $m_0$ is very large
and in this example is taken to be infinite.
We suppose that $O$ with total mass $m_1$ includes detector $D_o$
  at the distance   $d_d$ from c.m. and small (pointlike) aperture $\delta w$
  so that $D_0$
 internal wave function relative to $O$ C.M. is :
$\phi_d(x'_d)=\delta(x'_d-d_d) $.
Additional neutron detector $D_2$ is installed on the opposite chamber
wall and detect the neutrons which didn't interact with $D_o$.

 Due to independence which means the factorisation of $O$ and $n$ states
  according to the form (1)
  $n$  wave function in ORF $\psi'$ can be extracted from the
$O+n$  system wave function:
\begin {equation}  
  \psi(x_n,x_1)= \psi_1(x_1)\psi_n(x_n)=\Phi_c(X_c)\psi'_n(x_n-x_1)=
 \psi_1(\frac{m_1x_1+m_nx_n}{m_1+m_n})\psi_1(x_n-x_1-x_s)
\end {equation}
Function $\Phi$ describes the state of this system as the whole and
can't be found by no mesurement of $n$ in ORF. $\psi_n$ in fact put
constraint on its state and results in its correlation with $\psi'_n$.
Considering the collapse in different RF we note that
 $O$ and MRF observers will treat the same event unambiguously
as the $n$ detection  (or it flight through $D_0$). In
observer reference frame (ORF) it
reveals itself by the detection and amplification process in $D_o$
 initiated by $n$ absorption and recorded later in RD. For MRF
the collapse results from the nonobservation of neutron in a due time
in  $D_2$ - so called negative result experiment. So we conclude that
the signal in $ORF$ will have  the same relative probability as in
MRF. Such kind of the measurement means obviously the reduction of $\Phi(R)$
in MRF and  futher measurement of $\Phi$
will coincide with it. Because of it proceeding further we'll assume
always the results for the quantum ensemble of observers $O$ without
additional refering. As we
have no reason to assume that transition from MRF to ORF which we'll
call Q-transformation can transfer
pure states to mixed ones we must conclude that that this distribution
is defined by neutron wave function in ORF.
 After it we can regard $d_d $  as the variable which
describe in ORF the space coordinate distribution of neutrons which
constitute pointlike beam in MRF and that it  will be
 the same as the  $O$ space distribution in MRF.
Probability calculated in MRF for the neutron absorbtion in $D_o$ is :
\begin {equation}   
   P(d_d)=\sigma_n|\psi_d(x_s)|^2=\sigma_n|\psi'_n(d_d)|^2
\end {equation}
where $\sigma_n $ is the neutron absorbtion in $D_0$ cross section.
 It means that the result of measurement in ORF is also
described by QM Reduction postulate, i.e. that initial state during
 the measurement by RF detector evolve into the mixture of
measured observable eigenstates.
 It's especially important because in general case
it'll result in complicated correlations  in density matrix of final state.
This results demonstrate that if in MRF $O$ wave function have the average
$x$ dispersion $a_O$ then from the 'point of view' of
 observer $O$ any object localised in MRF is
smeared with the same RMS $a_O$.

After considering this qualitative picture we turn to the calculations
of more general situations. First we consider the case
 when $n$ distribution in MRF isn't pointlike
but is described by the arbitrary wave function $\psi_2(r_2-r_0,t)$
and $O$ distribution is given by $\psi_1(r_1-r_0,t)$.
We regard that all the objects evolution including $M_R$ are described by free
SE and in its description  we can neglect by the internal degrees of freedom
 of each object.
The total system Hilbert space is denoted as $H_s$, and $r_i$ are 3-vectors.
 We must find operator (functional) of $\psi_2$ unitary
transformation to ORF
$$
 \psi'_n(r_2-r_1,v)=\hat{U}_Q(\psi_1)\psi_2(r_2-r_0)
$$
where $v$ mean all other degrees of freedom it can depend.
If  to compare it with $\hat{U_G}$ - operator of inhomogenious
Galilean transformations which has 6 degrees of freedom ($\vec{v},\vec{a}$)
, $\hat {U_Q}$ will have formally infinite number of them defined by ORF state
vector $\psi_1$.
Due to it
 the $n$  wave function in ORF  $\psi'_n$ will acquires complicated
 entangled form which will be calculated transforming space coordinates
  $r_i$ to Jacoby coordinates  which transfer also our 3-particles
 system state $\cite{Schiff}$. We'll use 2 sets of them ${u^0},{u^1}$
 corresponding
to observers of MRF and ORF :
\begin {equation} 
u^0_1=r_2-r_0 , u^0_2=r_1-\frac{r_2m_2+r_0m_0}{m_0+m_2},
u^1_1=r_2-r_1, u^1_2=r_0-(r_1m_1+r_2m_2)/M_2
\end {equation}
  where
 $M_2=m_1+m_2$. We don't include $R_c$ in this sets and regard
it as additional system coordinate. Note that to each $u^i_j$
 corresponds Hilbert subspace
$H^i_j$ of $H_s$. Due to independence of $O$ and $n$ states
in MRF we calculate $n$ wave function in ORF which is in the same time is the
system wave function :

\begin {equation}   
      \psi'_n(u^1)=\psi'_s(u^1)=\hat{U}_Q(\psi_1(u^0))\psi(u^0_1)=
\psi_1(u^1_2+\frac{m_2u^1_1}{M_2})*
\psi_2(u^1_2-\frac{m_1u^1_1}{M_2})
\end {equation}
State vector $\psi'_n$ formally is the tensor product of two state
vectors which belong to subspaces $H^0_1,H^0_2$. This transformation
is equivalent to
relative cordinates axes rotation in $H_s$ defined by $u^0_i=c_{ij}u^1_j$
 relation as following from (5).
 Note that the result doesn't
 depend on $m_0$
,which as will be shown differ from the general situation
 when $n$ and $O$ states are correlated.

Now we'll account for dependence on $m_0$ which is also important
for the description of the system evolution. The most simple way
to perform it is to introduce additional (dummy) observer $O_A$ which
mass is very large and relative to which the new system wave
fuction $\psi_s(r_0,r_1,r_2)$ is defined. It satisfy to free Schrodinger
equation
for this 3 objects and can be factorised into $\Phi_c(R_c,t)\psi'_s(u^j,t)$.
This equation includes $m_0,m_1$ mass symmetrically, and if to
 extract from it c.m. movement accounted by $\Phi_c(R_c,t)$ we'll get
the equation for the relative movement $M_R ,O$ and $n$. The choice
of observer is equivalent of the choice of $u^j$ set.
 Note that by no measurements of $M_R,O$ or $n$ $\Phi_c$ can be defined by
either of MRF and ORF observers. Hence $\psi'_s$ is independent of $R_c$
 and defined by the internal evolution of this system.
The resulting equation for  $\psi'_n=\psi'_s$   the
 wave function in $u^1$ coordinates is :
\begin {equation} 
-\sum_{j=1}^{N-1}\frac{\Delta_{u}}{2\mu_j} \psi'_s(u^1,t)
   =i\frac{d\psi'_s}{dt}(u^1,t)
\end {equation}
where in this case N=3, $\mu_1=(m_1^{-1}+m_2^{-1})^{-1}$,
$\mu_2=(M_2^{-1}+m_0^{-1})^{-1}$,$\Delta_u$ is Laplasian of $u^1$.
Now we can annihilate dummy observer $O_A$ ,because obtained equation for
internal coordinates doesn't depend on its presence.

 In general equation (7) with $3(N-1)$ degrees of freedom
 $u_1,...,u_{N-1}$ for $N$ objects including
 observer $l$ ($l=1$ in a regarded case)
 can be regarded as the true QM evolution equation which account
the quantum movement of finite mass observer, neglected in Schrodinger
equation.
 C. m. movement of the  whole system is logically absent in it
,because it's equivalent to observer absolute movement which can't have
physical meaning. In this equation  which deals only with the
relative movement of the objects and observer we formally
can define any object $j$ of $N$ as observer , changing correspondently the
u coordinates set.  Formulaes  for $u$ set for  arbitrary $N$  and $l$
  can be deduced easely from Hamiltonian invariance under Jacoby
transformations $\cite{Schiff}$.
 When $m_l->\infty$ equation (7) transforms to Schrodinger equation
 for $N-1$ objects. Formally we can approximate $N$ to the number of the
objects in the universe, and regard $\psi'_s$ as its wave function
 ,demonstrating that no preferable RF in the universe exists.

In this framework the most general type of system Q-transformation is
when we'll have $N_o$ observers and $N-N_o$ 'particles' described by
the wave function $\psi'_l(u^l,t)$ of Jacoby coordinates in RF $l$.
 The transformation from
RF $l$ to $j$ is transformation of Jacoby $u^l$ set which change
state vector  $\psi'_j=\hat{U}_{jl}(\psi'_l)\psi'_l$.
Considering the group properties of Q-transformations we note that it's
finite group of the dimension $N_o$ ,for which existence of unit element is
obvious. The inverse element of $\hat{U}_{jl}$ is $\hat{U}_{lj}$. Any
arbitrary transformation is expressed in $l$ basis :
$\hat{U}_{km}=\hat{U}_{lm}\hat{U}_{kl}$.

 Now we turn to calculations for $N=3$ and $\psi'_n$ of (6), because
 they have simple physical interpretation.
 As follows from Schmidt theorem $\psi'_n$ can be decomposed at $t=t_0$
$\cite{Von}$
\begin {equation}   
      \psi'_n(u,t_0)=
  \int f(k)\varphi_1(k,u^1_1)\varphi_2(k,u^1_2)d^3k
\end {equation}
 where $\varphi_1$,$\varphi_2$ can be chosen so, that they form
ortonormal systems in $H^1_1$ ,$H^1_2$.
For time $t>t_0$ $\psi'_n$ will conserve initial entangled form
with the weights $f(k)$ and functions $\varphi'_1(k,u^1_1,t)
\varphi'_2(k,u^1_2,t)$
to be solutions of equations (7) with (6) as the initial values.
Probability to find some $u^1_1$ value in ORF is to be :
\begin {equation} 
   P(u^1_1,t) =\int |f(k)|^2|\varphi'_1(k,u^1_1,t)|^2d^3k
\end {equation}
This entanglement is the consequence of quantum correlations of
$n$ and $O$ with $M_R$ expressed by $\psi_2$ and $\psi_1$, which results in
correlations between $n$ and $M_R$, analoguous to the proton-electron
correlations in hydrogen atom.
This results don't mean that all the pure states in ORF are entangled
with $M_R$ state. For example if $n$ was emitted by $O$ itself then $\psi'_n$
 will
depend on $x_2-x_1$ only and will be disentangled from the $M_R$ state in ORF.
In this simpliest case we must take $N=2$ in (7) and as the result
$\psi'_n$ will depend on $u^1_1$ observable.
To calculate the change of state (8) due to the measurement of $u^1_1$
by ORF we calculate Density Operator of reduced state $\cite{Von}$ :
$$
   \rho_o(u^1_2,u'^1_2,t)=\int \psi_s(u^1_1,u^1_2,t)*
\psi^*_s(u^1_1,u'^1_2,t)du^1_1
$$
So the measurement transfers pure entangled state into partly mixed
state which stay to be pure for $u^1_2$ , i.e. for $r_1$ or $r_2$
observables.

As the example we'll take $\psi_1$ ,$\psi_2$
at $t_0$ to be gaussians with the RMS $a_1$,$a_2$ ,the distance between
their centers $\vec{d}$ and $n$ moving with the velocity $\vec{v}$.
 We perform the spectral decomposition by the Fourier transformation on $u_2$
and take momentum $p_2$ as the decomposition parameter $k$.

\begin {equation} 
   \psi'_n(u,t_0)= \int f(p_2)g(p_2,u_1)e^{ip_2u_2}d^3p_2
\end {equation}
It follows that (vector products can be easily identified):
 $$
f(p_2)=c_0*exp[a^2_1\sigma^{-2}(im_2vd+idp_2-a^2_2(m_2v+p_2)^2)] ,
$$
$$
   g(p_2,u_1)=exp[i(bp_2+\frac{vm_2a^2_2}{\sigma^2})u_1
  -\frac{(u_1-d)^2}{\sigma^2}],
$$
where $\sigma^2=a^2_1+a^2_2$ , $b=M^{-1}\sigma^{-2}(m_2a^2_2-m_1a_1^2)$
and $c_0$ is normalisation constant. Corresponding time dependent
functions are:
\begin{equation}
   \varphi'_1(p_2,u_1,t)=\frac{1}{(1+it'\tau^{-1})}
exp[\frac{-(u_1-d)^2-2iv_p(u_1-d)\tau-iv_p^2\tau t'}{\sigma^2(1+it'\tau^{-1})}]
\end {equation}
$$
   \varphi'_2(p_2,u_2,t)= exp(ip_2u_2-\frac{ip_2^2}{2\mu_2}t')
$$
where $t'=t-t_0$,
$\tau=\mu_1\sigma^2$,$v_p=(\mu_1\sigma^2)^{-1}(bp_2\sigma^2+vm_2a_2^2)$.
 Despite the complcated form of $\psi'_n$ it easy to show that the
probability distribution (9) of $n$ space coordinate in ORF $r'_2=u_1$ will
 have gaussian form with RMS
$$
     \sigma(t)=\sigma(1+\frac{t'^2}{\tau^2})^{\frac{1}{2}}
$$
Note that even if the initial state is factorised into $\psi_1*\psi_2$
the final state become entangled and depend on $m_0$.
 If $a_2$ is very large and so $\psi_2$
tend to a plane wave it transforms into the wave packet with average
 momentum $m_2v$ and constant $u_1$ probability distribution. Note that
if $m_0>>m_1,m_2$ wave functions (9) don't depend on $m_2$ at all,
conserving initial smearing $\sigma$. Note that even if $m_0,m_1 ->\infty$
it follows from (6) that one observer $O$ can be smeared in the RF
of the other and vice versa. It's scale is defined mainly by the initial
 function $\psi_d$ and
here we touch the problem of the QM classic limit,
where they are also involved. We don't plan to discuss
 it, stressing only the point that some observables which initially can
satisfy this limit at larger time can violate it as well.
We can tell that exact Q-symmetry is spontaneously broken by initial
cosmological conditions, when a short after Big Bang all particles
 were closely correlated in space, but this symmetry slowly restores
with the time.

 As follows directly from (11) if $a_1$  is
 currently very small Q-transformations will practically
  coincide with
Galilean space translations ,but they principally differ
when $a_1$ is large i.e. of
the order of macroscopic scale . Especially significant
it becomes when $\psi_1$ approach to the plane wave limit. Then
 all localised in MRF objects in ORF will be
 described by the wave packets with constant $u_1$ probability
distributions. In general qualitative distinction from Galilean
 transformations is that under Q-transformations space point isn't
 transformed into another point , but into a function on 3-space
which is reflected by  the  change of  $\delta(r)$ under  Q-transformations
 in (4) and (8).
  To find the relations between Q- and Galilean
 transformations $\hat{U_G}$ we can calculate systems state for several dummy
 observers $O_i$
 moving with the different velocities $\vec{v_i}$ and with space
shifts $\vec{a_i}$. But the extracted wave function of internal
coordinates $\psi'_s(u^1,t)$ doesn't depend on this parameters ;
, only $\Phi_c(R_c,t)$ which is nonessential for internal observers will
 depend on them according to standard
Galilean transformations $\cite{Schiff}$. So we conclude that
 $\hat{U_Q}$ represents the internal symmetry relative to Galilean
transformations and formally they commute.
Other properties which should have  this newly defined space-time like
 a continuity,differentiability  follows from this properties of the
 initial space-time
and  smoothness of wave functions used as the transformation kern
$\cite{Bacry}$. Note that Q-transformations conserve cannonical
commutation relations of $q,p$ operators, fundamental for any
Field Theory.

Relations and especially commutativity of Q-transformations and
Poincare group is to be analysed in the framework of
relativistic analog of equation (7) on which we work now.
The main idea is the same : to separate c. m. quantum movement and
relative movement of the system parts ,but to perform it in relativistic
case is much more complicated task $\cite{Wigner}$. In this case the
objects relative movement is defined by their invariant mass square $s$.
Omitting simple considerations analoguous to described above we get
 equation of the motion for $N=2$ :
\begin {equation} 
-[m_1^2+m_2^2+2m_1(m^2_2+\Delta_{u})^{\frac{1}{2}}]
^\frac{1}{2}\psi'(u^1_1,t) =i\frac{d\psi'}{dt}(u^1_1,t)
\end {equation}
The use of relativistic square roots operators is described
in $\cite{Shw}$. It's easy to see that in nonrelativistic limit
 this equation coincide with (7) after $M_2$ subtraction.
For $N>2$ this equation will acquire complicated form
 expressed through the hierarchy of subsytems
each characterised invariant mass operator depending of relative momentum
of susbsystem constituents.

In relativistic case we don't expect significant changes of our previous
conclusions, because there is
a preferable reference frame in this problem which have average velocity of
$O$ in which we can define $\psi'$. Moreover we consider in fact infrared
limit for macroscopic object, so the role of negative energy states must
be small. Note also that wave function  $\delta(x)$ under
 Lorentz transformation acquire space smearing analoguous to
our results $\cite {Wigner}$ .

Now we'll discuss briefly the technical feasibility of the experiments
with observers wave packets.
We'll consider set-up analoguous  to already described above GEO where no
 gravitation exists  in a free fall vacuum
chamber and the solid state detector-recorder initially rigidly fixed
 in solid fixator is released at $t=0$. After some time period its displacement
 is defined by the marker particles beam. As the detector-recorder system
in fact can be used any detector with the memory like photoemulsion which
have coordinate accuracy of the order .1 micron.
Especially attractive  seems to be plastic or crystall track detectors
which under electron microscopic scanning can in principle define  the
position of dislocation induced by particle track with the
accuracy up to several interatomic distances. The same order will have
the initial packet smearing $a_0$, because it's defined by the surface
effects between detector and fixator surfaces which extended to
 interatomic scale.
 If we suppose the mass of the detector
to be $10^{-10}$ gramm (the mass of emulsion grain which acts as the
elementary individual detector) and
 $a_0$ value $10^{-2} mk$ we get
the average centre mass deflection
of the order .1 mk for the exposition time $10^6$ sec i.e. about
one week.
Despite that the performance of such experiments will be technically
extremally difficult it's important nonethereless that
 no principal prohibitions  for them exist.

\section{\large{Concluding Remarks}}

  We've shown that extrapolation of QM laws on the
macroscopic objects demands to change the approach to the
space-time coordinate frames which was taken copiously from  Classical
Physics. It seems that QM permits the existence of the RF
 manifold, the transformations between which principally can't
be reduced to Galilean or Lorentz transformations.
This new global symmetry means that observer can't measure its own
spread in space, so as follows from Mach Principle it doesn't exist.
The physical meaning have only the spread of relative coordinates
of OS and some external object which can be measured by OS or other
observer.

Historically QM formulation started from defining the wave fuctions on
Euclidian 3-space $R^3$ wich constitute $H_s$. In alternative approach
we can regard $H_s$ as primordial states manifold $\cite{Bacry}$. Introducing
particular Hamiltonian results in assymmetry of $H_s$ which permit
 to define $R^3$ as a spectrum of the continuous observable $r$
which eigenstates are
 $|r_i>$. But as we've shown for several quantum objects this definition
become ambiguous and have several alternative solutions in $H_s$.

At first sight Q-transformation will violate locality principle,but
it's easy to see that it holds for each particular RF, despite that point in
one RF doesn't transforms into point in other RF.This is easy to see for
nonrelativistic potential $V(r_2-r_1)$ ,but we can expect it true also
in relativistic Field Theory.
 So we can suppose
the generalisation of locality principle for Q-transformations, which
yet must be formulated in a closed form.

In our work we demanded strictly that each RF must be quantum observer
i.e. to be able to measure state vector parameters. But we
should understand whether this ability is decisive property
 characterising RF ? In classical Physics this ability
doesn't influence the system principal dynamical properties. In QM at first
sight we can't
claim it true or false finally because we don't have the established theory
 of collapse.
 But it can be seen from our analysis that collapse is needed
in any RF only to measure the wave functions parameters at some $t$.
Alternatevely this parameters at any RF can be calculated given
initial experimental conditions without performing additional measurements.
It's quite reasonable to take that quantum states have objective meaning
and exist independently of
their measurability by particular observer,so this ability probably can't
 be decisive for this problem. It means that we can connect RF with the
system which doesn't include detectors ,which can weaken and simplify
our assumptions about RF. We can assume as most important for RF is
to reproduce space and time points ordering and record it
 ,as solid states like crystalls can.

 Author thanks B.Altshuler, V.Bykov ,V.Berezin and D.Finkelstein
 for fruitful discussions and valuable remarks.

\end{sloppypar}

\end{document}